\documentclass[aps,showpacs,amsmath,amssymb,nofootinbib,preprintnumbers]{revtex4}
\usepackage{graphicx}
\usepackage{color}

\def \be  {\begin{equation}}
\def \ee  {\end{equation}}
\def \bea {\begin{eqnarray}}
\def \eea {\end{eqnarray}}

\begin{document}

\preprint{ECTP-2010-08}

\title{Orders of Fermi-- and Plasma--Accelerations of Cosmic Rays}
\author{A.~Tawfik}
\email{drtawfik@mti.edu.eg}
\affiliation{Egyptian Center for Theoretical Physics (ECTP), MTI University,
Cairo-Egypt}
\author{A.~Saleh}
%\email{a.saleh@ectp.mti.edu.eg}
\affiliation{Egyptian Center for Theoretical Physics (ECTP), MTI University,
Cairo-Egypt}

\date{\today}

\begin{abstract}
The generic acceleration model for ultra high energy cosmic rays, which has been introduced in {\tt 1006.5708 [astro-ph.HE]}, suggests various types of electromagnetic interactions between cosmic charged particles and the different types of the plasma fields, which are assumed to have general configurations, spatially and temporally. The well-known Fermi acceleration mechanisms are also included in the model. Meanwhile Fermi mechanisms in non--relativistic limit adhere first-- and second--order of $\beta$, the ratio of particle's velocity relative to the velocity of the stellar magnetic cloud, in the plasma field sector, $\beta$ does not play any role, i.e. zero--order. In the relativistic limit, the orders of Fermi acceleration are only possible, when applying the corresponding conditions, either elastic scatterings or shock waves. Furthermore, it is found that the coefficients of $\beta$ are functions of the initial and final velocities and the characteristic Larmor radius.

\end{abstract}


%\Keyword: Cosmic Rays, Electromagnetic fields, Drift waves (plasma)
%\keyword: UHECRs, Varying Fields, Drift Energy
\maketitle

\section{Introduction}

Generic mechanisms for ultra high energy cosmic ray acceleration have been extended by including various particle field interactions with general configurations either spatially or temporally and/or both \cite{Tawfik:2010jh}. A special form for the $\mathbf{E}\times \mathbf{B}$ drift dates back to the eighties of last century, when Hillas introduced his well--known graph \cite{hillas1}. In Ref. \cite{Tawfik:2010jh}, the plasma field interactions are assumed to accumulate to the Fermi ones and vice verse. The latter are classified into two classes or orders, first-- and second--order  of the particle's velocity relative to the velocity of the magnetic cloud, $\beta$. Elastic and shock waves scatterings \cite{shockB} with the magnetic field represent the nature of the Fermi accelerations \cite{fermi49}. The energy gained according to Fermi explicitly depends on the relative particle's velocity $v$ with respect to the velocity of the rest frame $u$ of the magnetic cloud. The plasma field interactions that likely take into account different quantum numbers of the cosmic charged particle represent the core of the generic acceleration model \cite{Tawfik:2010jh}. 

In this letter, the orders of the averaged energy difference relative to the initial energy is studied according to generic model \cite{Tawfik:2010jh}, which has been applied in Ref. \cite{Tawfik:2010BB} as well. In the cloud's frame of reference, we first assume that the cosmic charged particle fulfills two conditions; resting in a certain region, where the electromagnetic (hereafter plasma) potential is nearly vanishing and entering the cloud with an initial velocity. The magnetic cloud is conjectured to adhere magnetic irregularities, shock waves and various types of electromagnetic fields.

\section{Orders of cosmic rays acceleration in non-relativistic limit}

As introduced in Ref. \cite{Tawfik:2010jh}, we first assume that a cosmic charged particle with mass $m$ at rest is being accelerated by static and uniform plasma fields. When elastic and shock waves (hereafter stochastic) scatterings are entirely excluded, then the final energy of the particle reads
\bea \label{eq:Ef1}
E_f &=& 2\, q\, \mathbf{v}\, \mathbf{B}\, \mathbf{R},  
\eea
where $\mathbf{B}$ is the magnetic field strength. $\mathbf{v}$ and $\mathbf{R}$ being the velocity and distance covered by the particle within the plasma fields, respectively. The particle gyrates and describes a helix/spiral curve around the magnetic field lines. At equilibrium, centrifugal and magnetic  forces are equal, so that $q\, \mathbf{B}=m\mathbf{v}/\mathbf{r}_l$, where $\mathbf{r}_l$ is the Larmor radius. Therefore, Eq. (\ref{eq:Ef1}) can be re--written as
\bea
E_f &=& 2\,m\, v^2\, \frac{\mathbf{R}}{\mathbf{r}_l}. 
\eea

If the initial velocity of the cosmic charged particle is finite, then the kinetic energy reads $m v_i^2/2$. The field interactions raise this energy value to $2\delta\,m\,v_f^2$, where $\delta=\mathbf{R}/\mathbf{r}_l$. In this case, the energy difference (hereafter gained energy) relative to the initial one is
\bea\label{eq:dEF}
\left\langle \frac{E_f-E_i}{E_i} \right\rangle &=& \mu -1,
\eea 
which apparently does not dependent on $\beta\equiv u/v_f$, where $v_f$ is the particle's final velocity and $u$ is the velocity of the magnetic cloud. In other words, $\beta$ is zero--order. On the other hand, the quantity $\mu$ offers a tool to relate initial and final velocities to each other
\bea\label{eqmu}
\mu &=& 4 \delta \frac{v_f^2}{v_i^2}.
\eea

Now, we can apply the generic acceleration model \cite{Tawfik:2010jh} that the cosmic charged particle undergoes also field interactions besides elastic and/or stochastic scatterings. In this case, the averaged gained energy reads
\bea \label{eqFclass2}
\left\langle \frac{E_f-E_i}{E_i} \right\rangle &=& (\mu -1)\beta^0 + 4\mu \left(\beta^1 + \beta^2\right).
\eea  
It is obvious that the quantity $<E_f-E_i>/E_i$ depends on three orders of $\beta$. Zero--order refers to the plasma interactions. Here, the coefficient is $\mu-1$, where $\mu$ is given by Eq. (\ref{eqmu}). The first-- and second--orders refer to Fermi mechanisms. The coefficient of each of them equals to $4\mu$. It is important to notice here that the Fermi mechanisms turn to be related to $\mu$, which in turn is depends on $\delta$. The latter gives the ratio of the distance covered by the cosmic charged particle inside the plasma field $\mathbf{R}$ to the characteristic Larmor radius $\mathbf{r}_l$.     
 
Equation (\ref{eqFclass2}) reflects two sectors. The first one, first term, is the plasma field sector, at finite $\mu$, as given in Eq. (\ref{eq:dEF}). The second one, is the Fermi sector (first-- and second--order) , in which $\mu$ appears as well, as a result of applying the generic acceleration model. Switching off the first sector means $\mu=1$. In this case, $\beta^0$ vanishes entirely and the coefficients of $\beta^1$ and $\beta^2$ turn to have the well--known value.

\section{Orders of cosmic rays acceleration in relativistic limit}

In the relativistic limit, the energy that the cosmic charged particle would gain through elastic and stochastic collisions with the magnetic cloud, can be obtained by applying Lorentz transformation between the particle's frame of reference (primed) and the cloud's one (unprimed), when the particle enters the magnetic cloud and between frame of reference (again primed) and the cloud's frame of reference (also unprimed) and the particle's one, when the leaves the magnetic cloud. We first assume that the particle enters the magnetic could with an angle $\theta_1$ and leaves it with another angle $\theta_2^{\prime}$. Both angles are constructed with the cloud's direction of motion. In the magnetic cloud, the cosmic charged particle is assumed to undergo different types of interactions. With reference to Fermi acceleration mechanisms, then
\bea
E_1^{\prime} &=&\gamma \, E_1 \left(1-\beta\,\cos\theta_1\right), \\
E_2 &=&\gamma \, E_2^{\prime}  \left(1+\beta\,\cos\theta_2^{\prime}\right), \\
E_1^{\prime} &=&E_2{\prime},
\eea
where $\gamma=1/\sqrt{1-\beta^2}$. In the relativistic limit, $\beta=v/c$. The final energy reads
\bea 
E_2 &=&\gamma^2 \, E_1 \left(1-\beta\,\cos\theta_1\right)\;  \left(1+\beta\,\cos\theta_2^{\prime}\right). \label{eqXFermiF}
\eea 
When applying the generic acceleration model \cite{Tawfik:2010jh}, it is assumed that the cosmic charged particle also interacts with the plasma fields. Therefore, the particle's final energy is 
\bea 
E_2 &=&\gamma^2 \, \left(E_1+2\,\delta\omega\,v_f^2\right) \left(1-\beta\,\cos\theta_1\right)\;  \left(1+\beta\,\cos\theta_2^{\prime}\right), \label{eqFermiF}
\eea 
where
\bea
v_f^2 &=& \frac{v^{\prime}\pm u}{1\pm u\,v^{\prime}/c^2},
\eea
and $\pm$ stands for the so--called head--on and head--tail collisions, respectively.
Equations (\ref{eqXFermiF}) and (\ref{eqFermiF}) can be used to calculate the fractional change in the energy $<(E_2-E_1)/E_1>$ 
\bea 
\left\langle\frac{E_2-E_1}{E_1}\right\rangle &=& \gamma^2\left[1+\mu-\beta\cos\theta_1-\mu\beta\cos\theta_1+ \beta\cos\theta_2^{\prime}+\mu\beta\cos\theta_2^{\prime}-\beta^2\cos\theta_1\,\cos\theta_2^{\prime}-\mu\beta^2\cos\theta_1\,\cos\theta_2^{\prime} \right]-1. \hspace*{5mm}
\eea
In order to estimate this expression, the averaged values of $\theta_1$ and $\theta_2^{\prime}$ have to be calculated. In fact, this is based on the differences between first-- and second--order Fermi acceleration mechanisms \cite{penth1}.

The second--order Fermi acceleration can be understood as elastic collisionless scattering with the magnetic irregularities. In this case
\bea
<\cos\theta_1> &=& - \frac{\beta}{3},\\
<\cos\theta_2^{\prime}> &=& 0,
\eea 
which leads to
\bea \label{dEFermi1}
\left\langle\frac{E_2-E_1}{E_1}\right\rangle &=& \mu\,\beta^0 + \left(\frac{\mu}{3}\ + \frac{4}{3}\right)\,\beta^2.
\eea
For Fermi shock waves acceleration
\bea
<\cos\theta_1> &=& - \frac{2}{3},\\
<\cos\theta_2^{\prime}> &=& \frac{2}{3},
\eea 
which leads to
\bea 
\left\langle\frac{E_2-E_1}{E_1}\right\rangle &=& \mu\,\beta^0 + \left(\frac{\mu}{3}\ + \frac{4}{3}\right)\,\beta + \left(\frac{13}{9} + \frac{4}{9}\mu\right)\beta^2, \label{dEFermi2} \\ 
   &\simeq& \mu\,\beta^0 + \left(\frac{\mu}{3}\ + \frac{4}{3}\right)\,\beta. \label{dEFermi2b}
\eea
For shock waves, it is assumed that $v<<c$  and therefore $\beta^2$ turns to be negligible.

In both expressions (\ref{dEFermi1}) and (\ref{dEFermi2b}), a zero--order term appears. It refers to plasma field interactions, in which $<(E_2-E_1)/E_1>$ directly depends on $\mu$. The coefficients of first-- and second--order Fermi are equal to $(\mu+4)/3$. This means that the generic acceleration model conserves the characteristics of both Fermi and plasma field interactions in the relativistic limit. When switching off the plasma field interactions, i.e. when $\mu$ vanishes, the well--known coefficients of $\beta^2$ and $\beta^1$ are generated.

\section{Discussion and Conclusions}

In non--relativistic limit, both orders appear in one equation, Eq. (\ref{eqFclass2}). The fractional of energy difference and initial energy depends only on $\mu$, which interestingly relates the final-- to the initial--velocities (or --energies). The quantity $\mu$ depends on $\delta$. The latter gives the ratio of the distance covered by the cosmic charged particle inside the plasma field $\mathbf{R}$ to the characteristic Larmor radius $\mathbf{r}_l$. So far, we conclude that both $\mu$ and $\delta$ characterizes very well the generic acceleration model \cite{Tawfik:2010jh}.

In the relativistic limit, the first term in rhs of both equations (\ref{dEFermi1}) and (\ref{dEFermi2b}) refers to the plasma field interactions. Here, the cloud's velocity $u$ or $\beta=v/c$ does not play any role. Therefore, zero--order $\beta$ can be multiplied with. The coefficients of second-- (Eq. (\ref{dEFermi1})) and first--order (Eq. (\ref{dEFermi2b})) Fermi are equal to $(\mu+4)/3$. It is obvious to conclude that the generic acceleration model adhere characteristics of both Fermi and plasma field interactions. To judge about this, let's assume vanishing plasma fields. In this case, $\mu$ vanishes as well and the well--known coefficients of $\beta^2$ and $\beta^1$ are secured.

Back to the non--relativistic limit, we find that all orders are combined in one line in Eq. (\ref{eqFclass2}). In this limit, $\beta=u/v$. As given in Eqs. (\ref{eq:dEF}) and (\ref{eqmu}), the gained energy in the plasma field sector directly depends on $\mu$ and $\delta$. Therefore, zero--order $\beta$ can be conjectured. According to the generic acceleration model, $\mu$ appears in the Fermi sector (first-- and second--order) as well. Switching off the plasma field sector means $\mu=1$. In this case, ${\cal O}(\beta^0)=0$ and the coefficients of $\beta^1$ and $\beta^2$ get the well--known value $4$.

We conclude that the generic acceleration model \cite{Tawfik:2010jh} offers an extension to the well--known Fermi acceleration mechanisms. In this extension, the quantum numbers of cosmic particles and nature of the electromagnetic fields of various astrophysical objects are taken into consideration. following the path of determining the order of Fermi acceleration, we introduce in this latter an estimation for the orders of $\beta$ in the generic acceleration model \cite{Tawfik:2010jh}. We find that that electromagnetic field interaction is characterized by zero--order $\beta$, in both relativistic and non--relativistic limits. The first-- and second-- order Fermi accelerations are reproducible, as well. On the other hand, the coefficients of $\beta$ reflect the configurations of affecting interactions and their types. Adjusting just one quantity, $\mu$, the switching between plasma and Fermi sectors turns to be possible.

\end{document}